# Compressive-Sensing-Enhanced First-Principles Calculation of Photoluminescence Spectra in Color Centers: A Comparison between Theory and Experiment for the G Center in Silicon

Jiongzhi Zheng[1], Lukasz Komza[2,3], Yihuang Xiong[1], Natalya Sheremetyeva[1], Changpeng Lin[4,5], Sinéad M. Griffin[3,6], Alp Sipahigil[2,3,7] and Geoffroy Hautier[1,*]

[1]*Thayer School of Engineering, Dartmouth College, Hanover, New Hampshire 03755, USA*

[2]*Department of Physics, University of California, Berkeley, Berkeley, California 94720, USA*

[3]*Materials Sciences Division, Lawrence Berkeley National Laboratory, Berkeley, California 94720, USA*

[4]*Theory and Simulation of Materials (THEOS), École Polytechnique Fédérale de Lausanne, CH-1015 Lausanne, Switzerland*

[5]*National Centre for Computational Design and Discovery of Novel Materials (MARVEL), École Polytechnique Fédérale de Lausanne, CH-1015 Lausanne, Switzerland*

[6]*Molecular Foundry Division, Lawrence Berkeley National Laboratory, Berkeley, CA 94720, USA*

[7]*Department of Electrical Engineering and Computer Sciences, University of California, Berkeley, Berkeley, California 94720, USA*

## Abstract

Photoluminescence (PL) spectra are a versatile tool for exploring the electronic and optical properties of quantum defect systems. In this work, we investigate the PL spectra of the G center in silicon by combining first-principles computations with a machine-learned compressive-sensing technique and experiment. We show that the compressive-sensing technique provides a speed up of approximately 20 times compared with the finite-displacement method with similar numerical accuracy. We compare theory and experiment and show good agreement for the historically proposed configuration B of the G center. In particular, we attribute the experimentally observed E-line of the G center to a local vibration mode mainly involving two substitutional C atoms and one interstitial Si atom. Our theoretical results also well reproduce and explain the experimental E-line energy shifts originating from the carbon isotopic effect. In addition, our results demonstrate that some highly anharmonic modes which are apparent in computed spectra could be absent experimentally because of their short lifetimes. Our work not only provides a deeper understanding of the G-center defect but also paves the way to accelerate the calculation of PL spectra for color centers.

---

* Geoffroy.Hautier@dartmouth.edu

## I. Introduction

Quantum technologies are poised to strongly disturb the way in which we compute, sense, and communicate [1]. Emerging material platforms such as quantum dots [2], trapped and ultracold atoms [3], topological qubits [4], superconducting circuits [5], and solid-state spins and defects [6,7] hold great promise for a wide range of applications in the field of quantum computing, communication, and sensing [1]. Color centers have attracted significant attention due to their potential applications in quantum information science, particularly in the context of quantum communication [8-12]. These quantum defects can be used as single-photon emitters or spin-photon interfaces, which play crucial roles in enabling efficient and reliable quantum communication. In recent years, quantum defects have been studied in a variety of hosts: diamond [6], SiC [7], gallium nitride (GaN) [13], aluminum nitride (AlN) [14], zinc oxide (ZnO) [15], silicon (Si) [16], and two-dimensional materials such as h-BN [17].

Photoluminescence (PL) spectra are central to the operation of color centers. The PL spectra of a color centers are typically characterized by a narrow zero-phonon line (ZPL) and a broad phonon sideband (PSB), which describe the process of an excited electron radiatively decaying to the ground state (GS) and the resulting structural relaxation [18,19]. The shape of the PL spectra provides valuable insights into the excitation energies and the interactions between excited electrons and lattice vibrations, i.e., electron–phonon couplings. The PL shape is also important in determining the efficiency of the defect in various quantum protocols as any emission out of the ZPL and into the PSB is a loss. PL spectra can now be computed using first-principles computations, providing an atomistic understanding of the vibrations at play in the defect during luminescence and potentially clarifying the exact atomic structure of the color center [16,18-21].

These features are essential as a purely experimental approach can face issues in sample preparation or the isolation of individual defect centers.

Among the growing number of hosts considered for quantum defects, silicon is especially appealing given the maturity of the semiconductor industry and the endless possibilities in terms of nanofabrication. Color centers in silicon have long been studied with a resurgence of interest since the advent of quantum technologies. Among the many known silicon color centers, the G center is appealing because of its emission in the telecom range and relative ease of fabrication, which has historically made it an appealing target for all-silicon lasers [22]. Recent experimental efforts have focused on controllable manufacturing of isolated G centers on the wafer-scale [23,24], with efficiency enhancements achieved through integration into photonic waveguides and cavities [25,26]. In addition, local deactivation of a single G center has also been demonstrated, highlighting a path towards post-fabrication fine tuning of quantum defects [27]. Despite these advances, the PL spectra have not yet been entirely characterized theoretically, including identification of the local vibrational mode.

In this work, we compute the PL spectra of the G center using first-principles computations and compare the results with experiments. Notably, we apply a machine-learning compressive-sensing (CS) approach rather than the traditional finite-difference method to compute the defect phonon spectra, resulting in a significant reduction in computational costs. CS is a powerful technique which enables recovery of sparse solutions from incomplete data [28]. By employing CS, we can concurrently identify significant harmonic terms and determine their corresponding values, enabling the efficient and accurate characterization of the vibrational properties of the system without requiring computation of all the individual force constants explicitly. We compare the CS technique to the finite-displacement method and use our results on the G center to confirm the expected configuration of the defect, identify local vibrational modes, and estimate a theoretical Debye–Waller value. Our work not only shines light on the G center but also demonstrates how CS can accelerate the computation of PL spectra of color centers.

## II. Methods

### 1. First-principles calculations

The *ab-initio* calculations within the framework of density functional theory (DFT) [29], utilizing the Vienna *Ab-initio* Simulation Package (VASP) [30], were performed for structural optimization of defect centers. The projector-augmented wave (PAW) [31] method was employed to treat the Si($3s^23p^2$), and C($2s^22p^2$) shells as valence states. The Perdew–Burke–Ernzerhof (PBE) [32] generalized gradient approximation (GGA) [33] and the hybrid Heyd–Scuseria–Ernzerhof (HSE06) functional [34] with a value of $\alpha$ = 0.25 for mixing parameters were used for the exchange-correlation functional. Herein, the G-center configurations A and B were embedded within a (3×3×3) supercell of silicon containing 217 atoms. The ionic positions were fully optimized using a plane-wave cutoff energy of 520 eV and a 1×1×1 gamma-point electronic *k*-point mesh, with a tight force convergence criterion of $10^{-3}$ eV·Å$^{-1}$ and a tight energy convergence criterion of $10^{-8}$ eV. We performed excited-state calculations for G-center configurations A and B by manually constraining the orbital occupations to excite one electron into an unoccupied band [16,35,36]. All the static DFT calculations of perturbed structures for forces were performed using both the PBE and hybrid HSE06 functionals, with 1×1×1 gamma-center electronic *k*-point mesh and energy convergence better than $10^{-8}$ eV.

### 2. Finite-displacement method and CS technique

The interatomic force constants (IFCs) can be regarded as derivatives of the potential energy with respect to atomic displacements from their equilibrium positions. The Taylor expansion of potential energy is formulated as follows:

$$E = E_0 + \sum_{lk}\sum_{\alpha} \Phi_\alpha(lk)u_\alpha(lk) + \frac{1}{2!}\sum_{ll'kk'}\sum_{\alpha\beta} \Phi_{\alpha\beta}(lk,l'k')u_\alpha(lk)u_\beta(l'k'), \qquad (1)$$

where $V_0$ is a constant, $\alpha$ and $\beta$ are the Cartesian indices, $u$ is the atomic displacement, and $l$ and $k$ are the labels of the atoms in each unit cell.

The force $\mathcal{F}_{\alpha,DFT}$ acting on $lk$ along direction $\alpha$ can be expressed by [37]

$$\mathcal{F}_{lk,\alpha,DFT} = -\partial E/\partial u_\alpha(lk) = -\frac{1}{2!}\sum_{l'k'}\sum_{\beta} \Phi_{\alpha\beta}(lk,l'k')u_\beta(l'k'). \qquad (2)$$

To extract harmonic IFCs using the finite-displacement method, the symmetry constraints such as point/space group symmetries, lattice periodicity, permutation, and rotational and translational invariance are enforced to search for independent IFCs and irreducible atoms [38]. We then generate the displaced patterns by moving one irreducible atom for all the independent IFCs and obtain accurate forces using DFT. Finally, using Eq. (2) and the finite-difference technique [39,40], we can obtain harmonic IFCs for the selected systems. Note that the systems with low-symmetry operations and large supercells require a large number of displaced patterns using the finite-displacement method [39,40].

To extract harmonic IFCs efficiently and accurately, the CS technique [28] can be employed to gather the physically significant terms of the harmonic IFCs using limited displacement–force datasets [41]. In contrast to the displacement–force datasets used in the finite-displacement method, we generate atomic structures from the fully relaxed supercell by applying a random-direction displacement of 0.01 Å for all atoms. Subsequently, a static DFT calculation is used to obtain accurate forces $\mathcal{F}_{DFT}$ according to the Hellmann–Feynman theorem for all the generated atomic structures.

The force $\mathcal{F}_{a,DFT}$ acting on the $a$th atom along direction $i$ can be expressed by [37]

$$\mathcal{F}_{a,DFT} = -\partial E/\partial u_a = -\sum_{\alpha \in A/S} \frac{1}{\alpha!} \sum_{\hat{s}\alpha \in S\alpha} \Gamma_{JI}(\hat{s}) \Phi_J(\alpha) \partial_{a0} u_I^{\hat{s}\alpha}, \qquad (3)$$

where $\boldsymbol{a}$ is a complex composite associated with the $a$th atom and direction $i$ $(x,y,z)$, $u$ is the displacement, $\alpha$ is a representative cluster comprised of two lattice sites for second-order force constants, $A/S$ is the orbit space, $\hat{s}$ is a symmetry operator under the crystal space group $S$, $\Phi$ is the interatomic force constant, and $\Gamma_{JI}$ is the matrix associated with the linear combination of symmetry operator $\hat{s}$ and the Cartesian indices $(I, J \equiv \{x, y, z\})$.

With the displacement–force datasets at hand, Eq. (3) can be rewritten as

$$\mathcal{F}_{DFT} = \mathbb{A}'\Phi, \qquad (4)$$

where $\mathcal{F}_{DFT}$ is a $3N_s N_p$ force vector from the DFT calculations. Here, $N_s$ represents the atomic number in the supercell and $N_p$ represents the number of the displaced patterns. $\mathbb{A}'$ is a sensing matrix of dimension $3N_s N_p \times N_\Phi$, and $N_\Phi$ is the number of IFCs. The sensing matrix $\mathbb{A}'$ is constructed from atomic displacements and can be written as

$$\mathbb{A}'(\boldsymbol{a}, \alpha I) = -\frac{1}{\alpha!} \sum_{\hat{s}\alpha \in S\alpha} \Gamma_{JI}(\hat{s}) \partial_a u_J^{\hat{s}\alpha}. \qquad (5)$$

Considering symmetry constraints including translational and rotational invariance [38], Eq. (4) can be further simplified as [37]

$$\mathcal{F}_{DFT} = \mathbb{A}'\mathbb{C}\boldsymbol{\phi} \equiv \mathbb{A}\boldsymbol{\phi}, \qquad (6)$$

where $\mathbb{C}$ is a $N_\Phi \times N_\phi$ matrix. Here, $N_\phi$ is the number of independent IFCs.

To this end, using CS [28] and the least absolute shrinkage and selection operator (LASSO) techniques [42], the sparse solution of harmonic FCs can be estimated by solving the following optimization problem:

$$\boldsymbol{\phi}_{LASSO}^{CS} = \arg\min_{\boldsymbol{\phi}} \|\boldsymbol{\mathcal{F}}_{DFT} - \mathbb{A}\boldsymbol{\phi}\|_2^2 + \lambda\|\boldsymbol{\phi}\|_1, \quad (7)$$

where the first term represents the conventional sum-of-squares $\ell^2$ of the fitting error for the training data, the second term is an additional $\ell^1$ penalty, and $\lambda$ is a crucial hyperparameter that allows fine tuning of the balance between the model sparsity and accuracy in regularization techniques and optimization algorithms. Smaller values of $\lambda$ lead to least-squares-like fitting, resulting in denser harmonic FCs, whereas larger $\lambda$ values produce very sparse underfitted harmonic FCs. The optimal $\lambda$, yielding the highest predictive accuracy for the model, can be estimated through five-fold cross-validation.

After obtaining the solution to the above optimization problem, the relative fitting error $\sigma$ can be defined as follows:

$$\sigma = \sqrt{\frac{\|\boldsymbol{\mathcal{F}}_{DFT} - \mathbb{A}\boldsymbol{\phi}\|_2^2}{\|\boldsymbol{\mathcal{F}}_{DFT}\|_2^2}}. \quad (8)$$

In this work, we performed IFC calculations using the finite-displacement method with the Phonopy package [39,43] and the CS technique (LASSO) using our in-house code Pheasy [44].

### 3. Vibronic coupling and PL line shape

Using Fermi's golden rule [45], the Frank–Condon approximation, and Huang–Rhys theory [46], the absolute luminescence intensity $I(\hbar\omega)$ is defined as [20,47]

$$I(\hbar\omega) = \frac{n_r \omega^3}{3\varepsilon_0 \pi c^3 \hbar} \left|\overrightarrow{\mu_{eg}}\right|^2 \sum_m \left|\langle \chi_{gm}|\chi_{eo}\rangle\right|^2 \delta(E_{ZPL} - E_{gm} - \hbar\omega). \quad (9)$$

Here, $n_r$ is the refractive index of the host material, $\hbar$ is the reduced Planck constant, $\mu_{eg}$ is the transition dipole moment, $\chi_{g(e)m}$ is the $m$th vibrational state of the ground (excited) electronic state, $\omega$ is the photon energy, and $E_{gm}$ is the energy of the vibrational state $\chi_{gm}$. Considering the complexity of determining the absolute luminescence intensity $I(\hbar\omega)$, the normalized luminescence intensity $L(\hbar\omega)$ will be adopted and can be expressed as

$$L(\hbar\omega) = C\omega^3 A(\hbar\omega), \tag{10}$$

where the spectral function $A(\hbar\omega)$ can be defined as

$$A(\hbar\omega) = \sum_m |\langle \chi_{gm}|\chi_{eo}\rangle|^2 \delta(E_{ZPL} - E_{gm} - \hbar\omega) \tag{11}$$

and $C$ is the normalization constant [19].

Using the generating function approach, Fourier transformation of the generating function $G(t)$ yields the line-shape function [47,48]

$$A(\hbar\omega_{eg} - \hbar\omega) = \frac{1}{2\pi}\int_{-\infty}^{\infty} dt\, G(t) e^{i\omega t - \lambda|t|}, \tag{12}$$

where

$$G(t) = e^{S(t) - S(0)} \tag{13}$$

and $\lambda$ is the broadening parameter of the ZPL. The time-domain electron–phonon spectral function $S(t) = \sum_k S_k e^{i\omega_k t}$ is evaluated as the Fourier transformation of the frequency-domain electron–phonon spectral function:

$$S(\hbar\omega) = \sum_k S_k \delta(\hbar\omega - \hbar\omega_k), \tag{14}$$

where the $\delta$ function is expressed by a Gaussian function, and $S_k$ is the partial Huang–Rhys factor and can be defined as

$$S_k = \frac{1}{2\hbar}\omega_k \Delta Q_k^2. \tag{15}$$

Here, $Q_k$ is the mass-weighted displacement associated with the $k$th phonon mode and given by [19]

$$\Delta Q_k = \sum_{\alpha=1}^{N}\sum_{i=x,y,z}\sqrt{m_\alpha}\Delta R_{\alpha i}e_{k,\alpha i}, \tag{16}$$

where $m_\alpha$ is the atomic mass of the $\alpha th$ atom in the supercell, $\Delta R_{\alpha i} = (R_{\alpha i})_e - (R_{\alpha i})_e$ is the displacement of the $\alpha th$ atom between the excited and ground-state equilibrium atomic position along the $i$th direction, and $e_{k,\alpha i}$ is a normalized eigenvector that describes the displacement of the $\alpha th$ atom along the $i$th direction in the $k$th phonon mode.

The localization of a phonon mode can be described using the inverse participation ratio (IPR) [19,49]:

$$IPR_k = \frac{1}{\sum_\alpha |e_{k,\alpha}|^2}. \tag{17}$$

$IPR_k = 1$ indicates that only one atom vibrates in the N-atom supercell with highly localized character for the $k$th phonon mode. In contrast, when $IPR_k = N$, it indicates that all atoms in the supercell vibrate and that the corresponding mode is highly delocalized. In this work, phonon eigenstates, i.e., phonon eigenvalues and eigenvectors, were calculated using the **Phonopy** package [39,43], and the PL spectra calculations were performed using the **Pyphotonics** package [50].

### 4. Experiment

G centers were created in the 220nm device layer of a silicon-on-insulator wafer. The sample was implanted with 12C at 36 keV with a fluence of $10^{13}$ cm$^{-2}$ followed by a 20-s rapid thermal anneal at 1000°C in a nitrogen environment. Subsequent proton irradiation at 200 keV with a fluence of $10^{13}$ cm$^{-2}$ substantially increased the G-center yield. The sample was cooled down to 3.8 K in a cryostat, where PL spectra were collected through a 0.42 NA objective. The G centers were excited non-resonantly using 10 µW of 635-nm light, and the emission was collected through the same objective. The emission spectra were generated by an InGaAs spectrometer with a quantum efficiency of approximately 50%, where spectra were acquired over 10 s. Additional experimental details can be found in the literature [26].

## III. Results and Discussion

### 1. Atomic and electronic structures

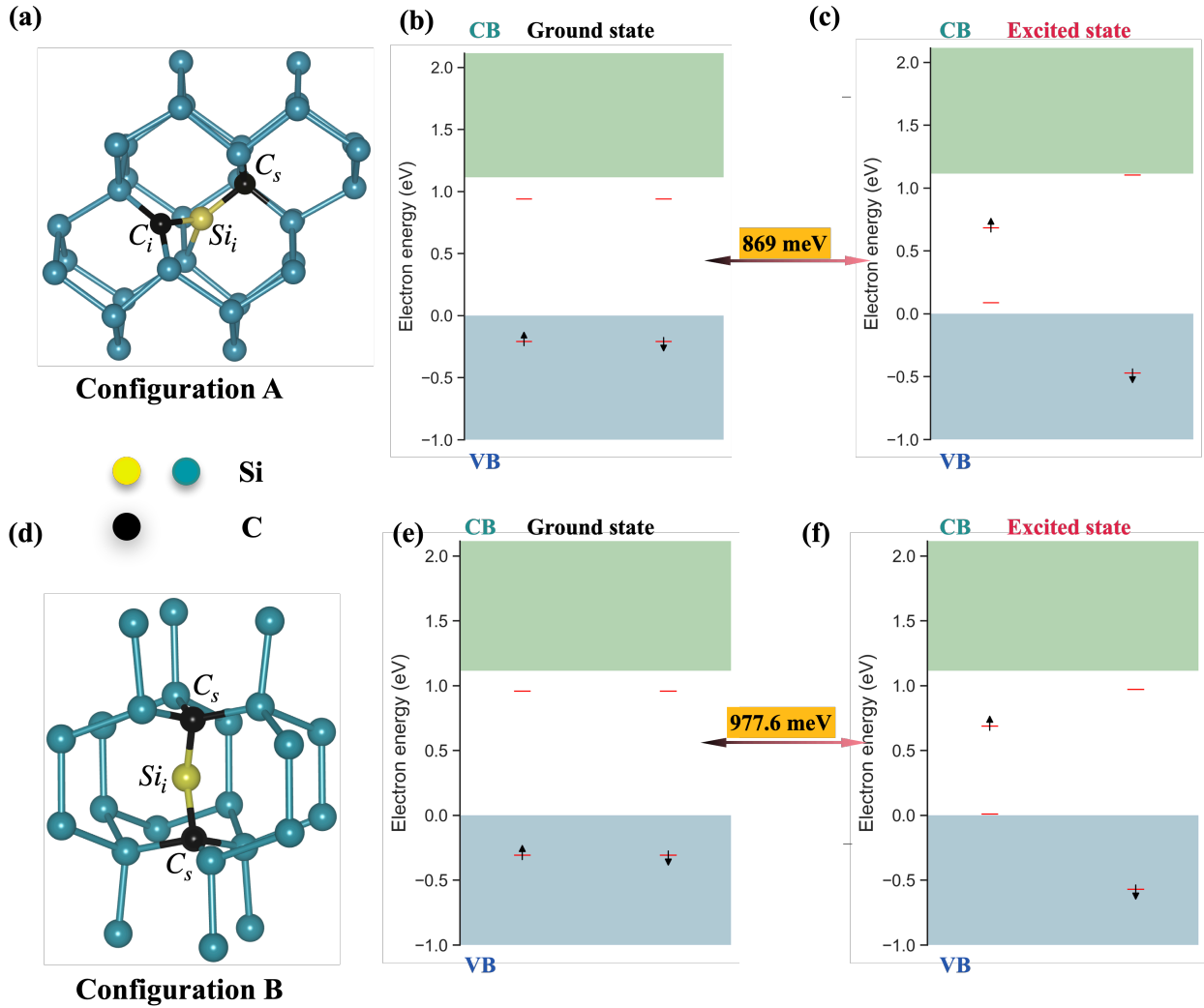

FIG. 1. (a) Atomic structure of defect center of configuration A, consisting of a carbon–silicon split-interstitial pair and a neighboring substitutional carbon atom. The yellow, green, and black spheres represent the interstitial Si, host Si, and C atoms, respectively. (b) GS electronic structure of configuration A. (c) ES electronic structure of configuration A. (d) Atomic structure of defect center of configuration B, which features two substitutional carbon atoms and an interstitial silicon atom (e) GS electronic structure of configuration B. (f) ES electronic structure of configuration B.

The lattice constant and electronic band gap of the silicon unit cell were first calculated using the PBE and HSE06 functionals. Negligible differences were found in the lattice constants of the silicon (Si) unit cell calculated using these two functionals: 5.47 Å for PBE and 5.43 Å for HSE06. Conversely, the difference in the calculated electronic band gap of Si was significant, with values of 0.611 eV for PBE and 1.11 eV for HSE06. Importantly, the lattice constant and band gap computed using the HSE06 functional well align with experimental results [51,52], and the HSE06 functional was thus selected for electronic-structure calculations on the defect centers in silicon.

Historically, two configurations have been proposed for the G center: configuration A and B both with a $C_{1h}$ point group symmetry [53,54]. Configuration A consists of a C–Si split-interstitial pair and a neighboring substitutional C atom ($C_iSi_i - C_s$) [see Fig. 1(a)]. Configuration B features two substitutional C atoms and one interstitial Si atom between two C atoms ($C_s - Si_i - C_s$) [see Fig. 1 (d)]. In this work, to model the defect centers, both configurations A and B were embedded in a $3 \times 3 \times 3$ supercell of Si, each containing 217 atoms, and the electronic structure and optical excitation for both configurations were investigated. The electronic structure calculations were conducted using the HSE06 functional, known for its proven accuracy in predicting the electronic structure of various defect centers, as discussed above [21,55]. The current study only concentrates on localized-to-localized transitions, where electronic transitions involve electrons/holes in localized states [26]. In the ground state (GS), the defect levels in G-center configurations A and B exhibit similar behavior. Figure 1(b) illustrates the presence of two highly localized defect levels in both G-center configurations A and B. Specifically, the occupied level $a'$ lies below the valence band maximum, whereas the empty level $a''$ is situated within the band gap. These observations are consistent with our previous calculations and findings from other studies [26]. Using the delta self-consistent field method (ΔSCF) [35,56], we excited an electron from the occupied level $a'$ to the empty level $a''$ [see Fig. 1(c)]. In the excited state (ES), we note that the empty level $a'$ is lifted above the valence-band edge, a phenomenon also observed in the negatively charged splitting vacancy in diamond [35,57]. The ZPL can be determined by calculating the total energy difference between the GS and ES of defect centers. The computed ZPL energies for the intra-defect transitions are 869 and 978 meV for configurations A and B, respectively, as depicted in Figure 1(c). The computed ZPL energy of configuration B demonstrates favorable agreement with our previous calculation [58], and other theoretical studies [16,59], as well as the experimentally measured value for the G center, i.e., 969 meV [60-62]

## 2. Fitting parameters of IFCs and phonon density of states

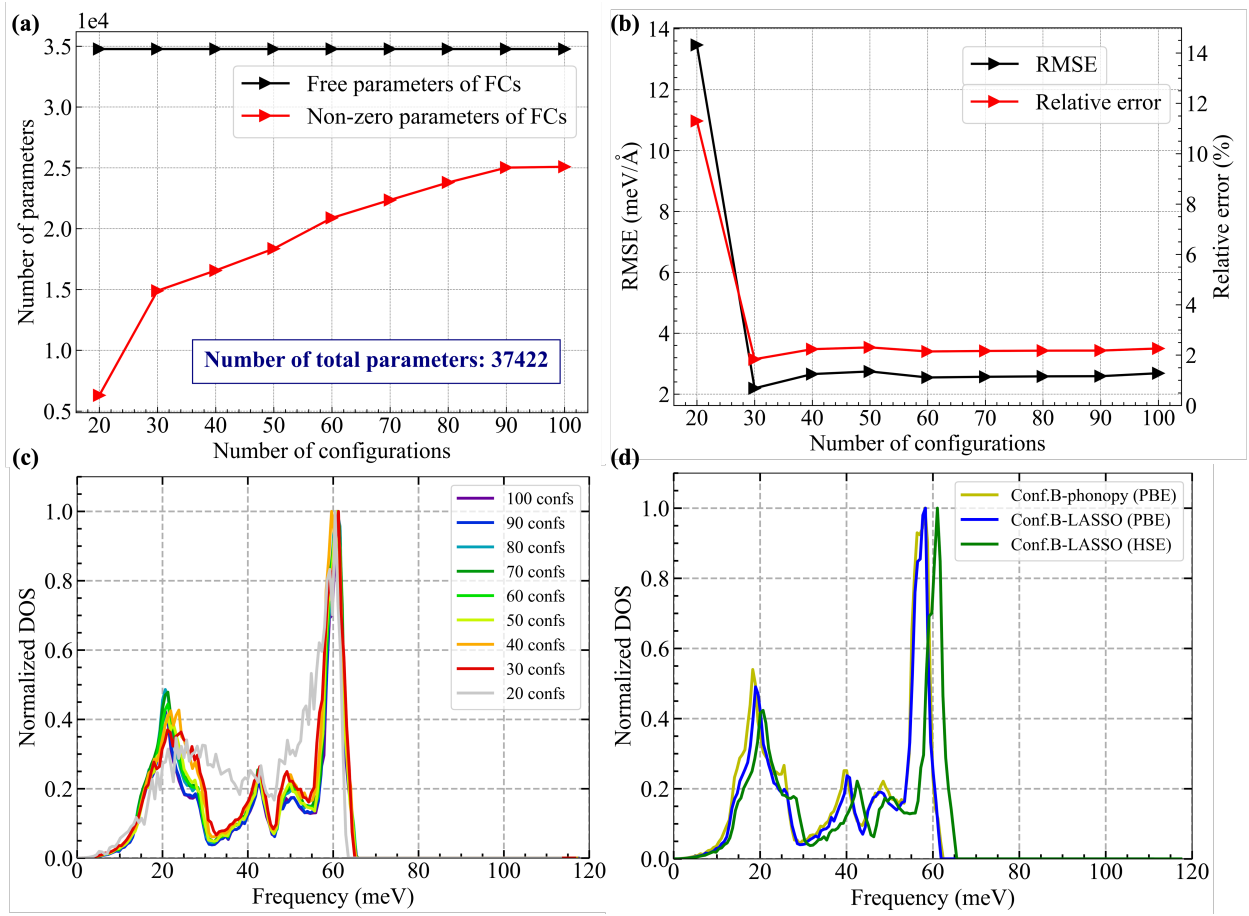

FIG. 2. (a) The number of free and non-zero IFCs as a function of the number of configurations. By applying symmetry constraints [63], the total number of IFCs decreases from 37,422 to ~35,000. (b) RMSE and relative fitting error with respect to the number of configurations. (c) Comparison of phonon DOS among various datasets. (d) Calculated phonon DOS using the finite-displacement approach and LASSO for configurations A and B, respectively. Because of the substantial computational consumption (688 configurations), we did not calculate the phonon DOS using the finite-displacement method [39,40] with the HSE06 functional.

With the optimized GS structure available, we next examined the key components, phonon eigenvectors and eigenvalues, used to determine the Huang–Rhys factor and PL line shape [19]. The finite-displacement method [39,40] has conventionally been employed to extract IFCs and compute phonons for both pristine crystals and defect centers [20,40,43,55,64]. Recently, the CS technique [28] along with the LASSO method [37,42] was employed to efficiently and accurately extract IFCs [41,64]. The low-symmetry operations ($C_{1h}$) combined with the large supercell of G-center configuration B led to a significant total of 35,000 free harmonic IFCs, necessitating 688 configurations for IFC calculations using the finite-displacement approach [see Fig. 2(a)]. Thus, we applied the CS technique to reduce the configurations and efficiently extract harmonic IFCs of G-center configuration B. Here, we examine the convergence of IFCs as the number of training datasets (configurations) is increased for G-center configuration B. Using the CS technique, one can substantially diminish the less significant terms in IFCs, thereby concurrently decreasing the requisite number of displaced configurations [see Fig. 2(a)]. In the case of 20 configurations, appropriately 29,000 IFCs are considered as physically unimportant terms and will be set to 0 in the CS technique. Even for 100 configurations, approximately 10,000 IFCs are still regarded as physically unimportant terms. To obtain converged results for IFC calculations using the CS technique, we calculated the relative error and root mean squared error (RMSE) of the LASSO method with respect to the dataset size. In Fig. 2(b), we observe that the errors achieve good convergence when the dataset size is larger than 20 configurations. Therefore, small relative fitting error ($\leq 2.5\%$) and RMSE ($\leq 3$ meV/Å) values are used as criteria for extracting harmonic IFCs for G-center configuration B, i.e., using 30 configurations. This observation underscores the advancement and efficiency of the LASSO method in extracting harmonic IFCs for defect centers. More specifically, compared with the finite-displacement method, the CS technique results in a speedup factor of up to ~23 for phonon calculation of G-center configuration B [see the compared details in Tab. S1 in the Supplemental Materials [65]].

Next, we investigated the convergence of the phonon density of states (DOS) in terms of dataset size, as depicted in Fig. 2(c). Specifically, the calculated DOS from 20 configurations exhibits a notable deviation from the others. Nevertheless, datasets exceeding 20 configurations exhibit a converged form for the DOS of the G center [see Fig. 2(b)]. This convergence is further evident in the phonon dispersions obtained from various datasets, as depicted in Figs. S1(a–i) and S2(a–i) of the Supplementary Materials [65]. To further validate the accuracy of the LASSO method [42], we calculated the phonon DOS for G-center configuration B using both the finite-displacement (*Phonopy* [39,43]) method with PBE and the LASSO method with PBE and HSE06 functionals [see Fig. 2(d)]. Unsurprisingly, the phonon DOS results calculated using the finite-displacement method (*Phonopy*-PBE) are quantitatively very similar to those obtained using the LASSO technique (PBE) [see Fig. 2(d)]. A strong correspondence in phonon dispersions is evident when comparing results obtained using the finite-displacement method and CS technique, as illustrated in Figs. S3(a–d) in the Supplementary Materials [65]. The calculated phonon energies using the HSE06 functional are generally higher than those obtained using PBE, which is attributed to the smaller lattice constant obtained using the HSE06 functional [see Fig. 2(d)]. Our results demonstrate the ability of the CS technique to calculate phonon DOS efficiently and accurately for defect centers with large supercells and low symmetry.

### 3. Calculated PL spectra

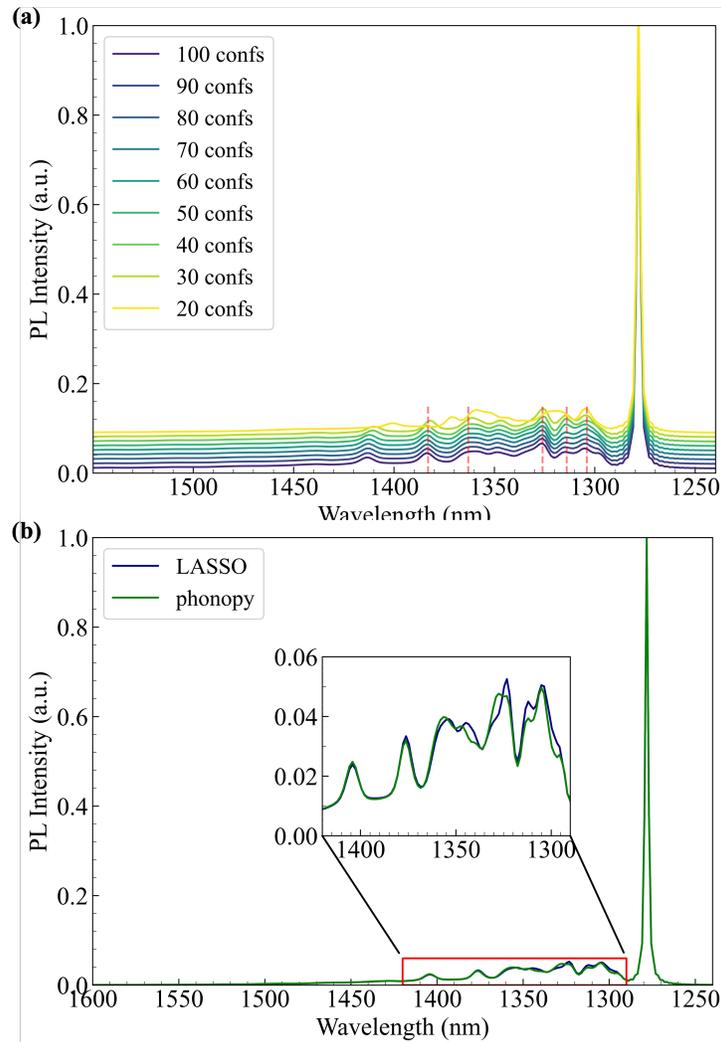

FIG. 3. (a) Calculated PL line shape using HSE06 functional and various datasets. (b) Comparison of PL line shapes obtained from Phonopy and LASSO technique.

Having obtained the optimized GS and ES structures along with $\Gamma-$point phonons, we next illustrate the convergence of the PL line shapes computed using both the PBE and HSE06 functionals, varying the dataset size from 20 to 100 configurations [see Figs. 3(a) and S4 in the Supplemental Materials [65]]. In line with the converged form of the DOS [see Fig. 2(c)] and phonon dispersions [see Figs. S2 and S3 in the Supplemental Materials [65]], the PL line shape obtained with 30 configurations exhibits satisfactory convergence. To validate the precision of the PL line shape acquired through the LASSO technique, we compare it with that calculated using the finite-displacement (*Phonopy*) method. The findings are depicted in Fig. 3(b), revealing a minimal and negligible difference between the outcomes of the two methods. These results demonstrate that 30 configurations are sufficient to obtain reliable and accurate results in the calculation of the PL line shape for the G-center configurations within the supercell size employed in this study.

## 4. Comparison between theory and experiment and identification of local vibration modes

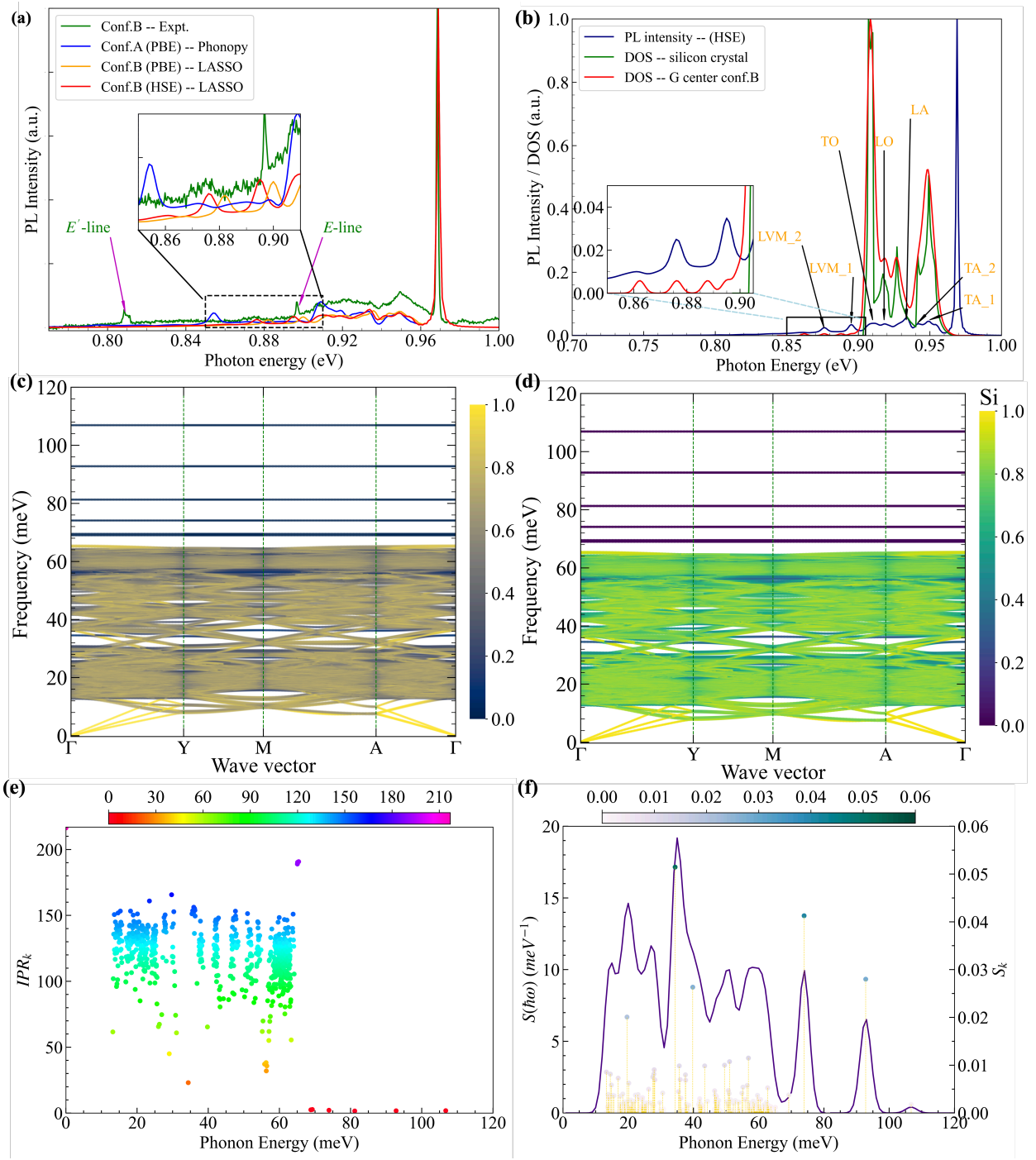

FIG. 4. (a) Calculated PL line shape of configurations A and B using finite-displacement method and LASSO technique compared with experimental results. (b) Calculated PL spectra and DOS of bulk silicon and G-center configuration B. (c) Calculated participation ratio projected onto the phonon bands along high-symmetry paths. (d) Calculated atomic participation ratio projected onto the phonon dispersions along high-symmetry paths. (e) Spectral

density of phonon–electron coupling and partial Huang–Rhys factors. (f) Calculated inverse participation ratio at Γ-point in Brillouin zone.

Fig. 4(a) presents a comparison between our computational PL results and the experimental data. Here, to facilitate a meaningful comparison, all the theoretical PL line shapes were adjusted to match the experimental ZPL [19]. From Fig. 4(a), we directly observe a significant discrepancy between the computed spectra and experiment for configuration A. The characteristic E-line peak (corresponding to local vibrational mode at 896.5 meV) observed in experiments is absent in configuration A [61,62,66]. Configuration A shows a local vibration mode (LVM) at 854 meV that is absent from experiment. Due to the observed discrepancies in both the ZPL and PL line shape between configuration A and the experimental G center, it can be concluded that the experimental G center cannot be in configuration A. This finding aligns with the conclusions from Gali *et al.* [16]. In contrast, the computed PL line shape of configuration B accurately captures the characteristic E-line mode, as highlighted in Fig. 4(a). Additionally, we observe a striking consistency in the results obtained from both the PBE and HSE06 functionals, as previously observed for the NV center [19]. This finding suggests that the PBE, while being inaccurate for the ZPL, accurately captures the phonons, successfully reproducing the PSB of G-center configuration B.

Although there is generally favorable agreement between the theoretical and experimental PL line shapes for configuration B, subtle discrepancies are observed [see Fig. 4(a)]. Specifically, the $E'$-line peak was identified in both our experiment and in previous studies at approximately 0.81–0.825 eV [61,62,66]. Surprisingly, this distinctive $E'$-line peak is absent in our theoretical results [see Fig. 4(a)], creating an inconsistency with the assigned peak found in the literature [59]. This inconsistency can be attributed to the limitations of the current one-phonon Huang–Rhys theory, neglecting the effect of multi-phonon processes [46,67]. To observe the $E'$-line peak in theoretical calculations, a phonon energy of ~144 $meV$ is needed. However, the highest phonon energies computed using PBE and HSE06 functionals are 120 and 104 $meV$, respectively. To observe this peak in theory, the application of the multi-phonon Huang–Rhys theory may be necessary [67]. An additional discrepancy between theory and experiment is the presence of a PL peak with a photon energy of 876 meV in theory that is absent in experiment (designated as the E-satellite peak in this work) [61,62,66]. We attribute the absence of this peak in experiment to the anharmonicity in this local vibration mode and the short lifetime that precludes its experimental measurement (see Section 5).

In Fig. 4(b), the PL line shape is overlaid on both the bulk-silicon phonon DOS and the DOS within the supercell containing a defect. The broad feature observed experimentally below the ZPL up to 904 meV primarily originates from bulk silicon phonons, incorporating characteristics of the transverse optical (TO), longitudinal optical (LO), longitudinal acoustic (LA), and transverse acoustic (TA) modes [62]. Beyond the energy range of bulk phonons (64 meV), two defect-related LVMs emerge: LVM1 and LVM2. To better understand the influence of phonon modes on the emergence of the PSB, we delve into the analysis of lattice vibrations linked to the localization in the G-center configuration B. Therefore, we proceed to calculate the participation ratio (PR) [68] projected onto the phonon dispersions and inverse participation ratio (IPR) [19,49] of configuration B [see Figs. 4(c–e)]. The phonon dispersion in configuration B differs from that of bulk silicon, which typically exhibits highly delocalized modes [see Fig. S5 in the Supplemental Materials [65]]. However, in configuration B, specific phonon branches show a small PR across the Brillouin zone, which are identified as quasi-localized or localized modes [see Figs. 4(c) and (e)] and are dominated by interstitial Si or/and C atoms [see Fig. 4(d) and 5(a–d)]. We further calculate the partial Huang–Rhys factor along with the spectral function of electron–phonon coupling, as shown in Fig. 4(f). Our calculations suggest that the PSB can be produced by a combination of approximately six phonon replicas, characterized by large Huang–Rhys factors and quasi-localized and localized nature. Here, we present two typical quasi-localized modes (QLM_1 and QLM_2) and two localized vibrational modes (LVM_1 and LVM_2) in Figs. 5(a–d). Despite the perturbation introduced by defect-induced phonon modes, two prominent modes with frequencies of 34.5 meV (QLM_2) and 73.9 meV (LVM_1) exhibit relatively small partial Huang–Rhys factors of 0.051 and 0.041, respectively [see Figs. 4(f) and 5(b–c)]. These values account for only 8.4% and 6.8% of the total Huang–Rhys factor, respectively. This observation suggests that both bulk and defect-induced phonons contribute to the PBS, with the primary influence on phonon–electron couplings in configuration B stemming from bulk phonons.

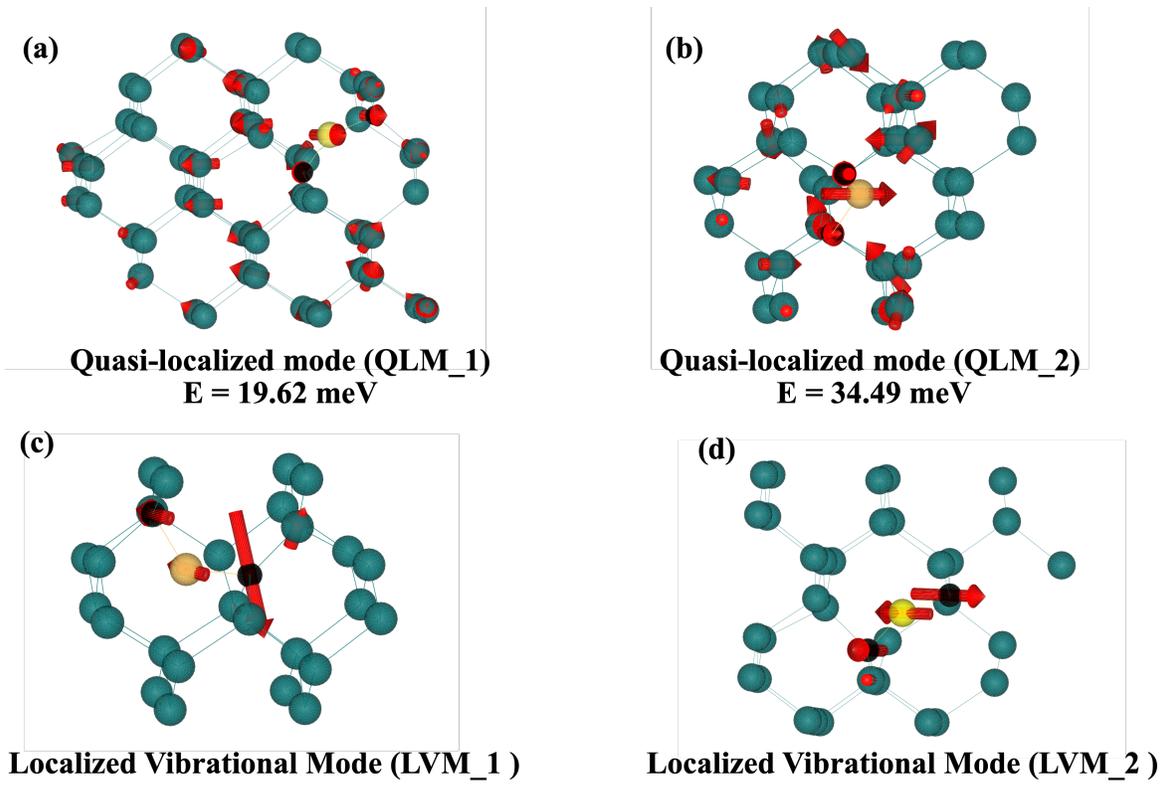

FIG. 5. Phonon normal mode for G-center configuration B. (a) Quasi-localized mode with energy of 19.62 meV. (b) Quasi-localized mode with energy of 34.49 meV. (c) Localized vibrational mode (LVM_1) associated with E-line peak. (d) Localized vibrational mode (LVM_2) associated with E-satellite peak. The yellow, green, and black spheres represent the interstitial Si, host Si, and C atoms, respectively. The red arrows indicate the mass-weighted displacements of each atom, with the magnitudes amplified by a factor of 3 for enhanced clarity.

## 5. Potential energy surfaces and isotopic effect

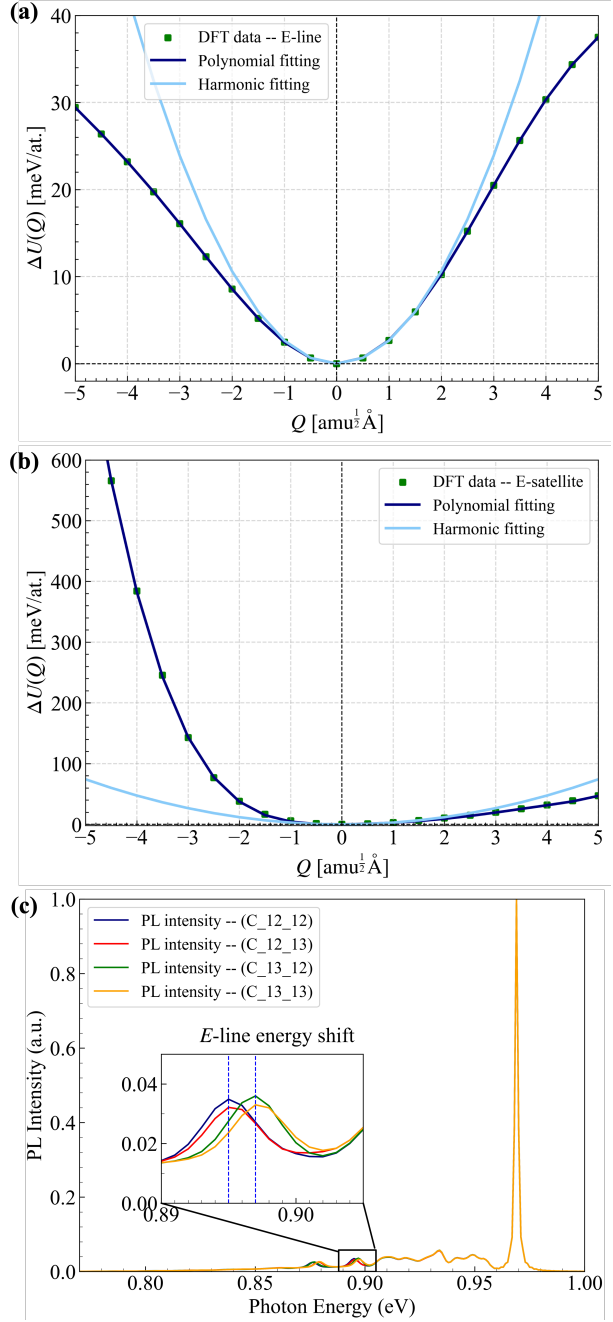

FIG. 6. (a) Calculated potential energy surface of phonon mode at gamma point associated with E-line peak. (b) Calculated potential energy surface of phonon mode at Γ-point associated with E-satellite peak. (c) Isotopic effect on PL line shape.

Despite the progress made in understanding the PL line shape of configuration B, the E-satellite peak is absent from experiments [61,62,66]. We note that the current theory does not explicitly consider a phonon lifetime and assume a similar broadening set-up arbitrarily for all phonon modes. Anharmonicity could affect lifetime and make some modes broader than others to the point that they will be more difficult to observe. To study the anharmonicity of phonon modes, we proceed to calculate the potential energy surface for the vibrational modes activating the E-line and E-satellite peaks, as illustrated in Figs. 5(c–d) and 6(a–b), respectively. Clearly, the potential energy surfaces of the vibrational modes corresponding to the E-line and E-satellite peaks feature completely different shapes [see Figs. 6(a–b)]. In particular, the phonon mode associated with the E-line peak exhibits a harmonic behavior. In contrast, the potential energy surface of the mode associated with the E-satellite peak is highly anharmonic, and attempts to fit it with a harmonic model have proven unsuccessful. Indeed, phonon modes with strong anharmonicity are known to exhibit large linewidths. The anharmonicity in the potential energy surface leads to non-linear and complex vibrational behavior, causing broadening of the phonon modes in the spectral domain [69]. Based on the observation of the highly anharmonic potential energy surface and the large linewidth associated with the phonon mode responsible for the E-satellite peak, we can attribute the disappearance of the E-satellite peak in experiments to the significant broadening of the PSB. We proceed with the analysis of the isotopic effect on the peak of the PL line shape, as illustrated in Fig. 6(c). We specifically focus on the isotope effect originating from the carbon atom, which serves as further evidence to aid in identifying the specific configuration of the G-center defect in experiment [61]. As previously mentioned, the lattice vibrations of the phonon mode associated with the E-line peak are primarily contributed by the carbon atoms of the defect center [see Figs. 4(c–d)] and 5(c–d). This observation emphasizes the significance of the role of the carbon atom on the phonon behavior and, consequently, the PL characteristics of the G-center defect. A G center made with $^{13}C$ implantation shows a splitting of the E-line (LVM1) peak into two peaks. We simulated the addition of $^{13}C$ by changing the mass of the carbon atom in our computation and found an agreement with this experimental finding with a splitting into two peaks as well. Specifically, the theoretically predicted energy shift of the E-line peak due to the isotopic effect is 2 meV, closely matching the experimental value of 1.95 meV [61]. We note that early in the study of the G center [61], it was argued that the defect should consist of only one carbon and one silicon atom, primarily based on the observation that only one additional PL line-shape peak was detected in experiment following the implantation of $^{13}C$. Our computations confirm that a model of two carbon and one silicon interstitial, such as configuration B, is however consistent with the experiment.

## 6. Huang–Rhys and Debye–Waller factors of defect centers

**Tab. 1. Huang–Rhys and Debye–Waller factors**

|  | PBE functional (Conf. B, this work) | HSE functional (Conf. B, this work) | Theory Ivanov *et al.* [59] (Conf. B) | Expt.[23-25] (Conf. B) | Expt.(this work) |
|---|---|---|---|---|---|
| $\Delta R$ (Å) | 0.076 | 0.076 | × | × | × |
| $\Delta Q$ (amu$^{1/2}$ Å) | 0.391 | 0.391 | × | × | × |
| Huang-Rhys factor ($S$) | 0.573 | 0.605 | 0.127 | × | × |
| Debye-Waller factor | 56.40 % | 54.63 % | 82.9 % | 15, 20, 30 % | 16 % |

Finally, using the optimized GS and ES structures, as well as the phonons, allows us to calculate not only the PL line shape but also the Huang–Rhys factor, as illustrated in Tab. 1. The Huang–Rhys factor is a crucial parameter that quantifies the strength of electron–phonon coupling in a defect center, representing the average number of phonons emitted during an optical transition [19]. As a result, the weight of the ZPL in the PL spectrum, commonly known as the Debye–Waller factor (DWF), is defined as $DWF = e^{-S}$. We employ both PBE and HSE06 phonons to calculate the Huang–Rhys and Debye–Waller factors for the G-center defect [see Tab. 1]. The Debye–Waller factor of G-center configuration B obtained using the HSE06 phonons is 54.63%, notably lower than the calculated value reported by Ivanov *et al.* (82.9%) [59]. The previously measured experimental values fall within the range of 15% to 30% [23-25], and our measured result is 16%. The comparison between the experimentally measured and theoretically predicted results is not as good as that observed for the NV center [19]. The discrepancy between the theoretically predicted and experimentally measured Debye–Waller factor can be attributed to the following factors: ($i$) the limitation of current one-phonon Huang–Rhys theory, as multi-phonon processes may be required [67]; ($ii$) the limitation of the equal-mode approximation used in the Huang–Rhys theory; ($iii$) the finite supercell size effect [19]; ($iv$) the difficulty in isolating a single defect center in experiments; and ($v$) the quality and purity of the experimental samples.

## IV. Conclusions

We investigated the vibrational spectrum of the G center in bulk silicon using a combination of first-principles DFT calculations and experimental measurements. Using a first-principles-based machine-learning CS technique, we were able to speed up the calculation of the vibrational spectrum while maintaining accuracy. We have demonstrated a significant speedup factor of ~20 for the CS technique compared with the traditional finite-displacement approach. This advancement renders PL spectra calculations feasible for quantum defect centers using hybrid functionals or the use of these approaches in high-throughput schemes [58,70,71].

In addition, we used our theoretical results on the G center to confirm the expected configuration of the defect, clearly demonstrating that the G center cannot be in historically proposed configuration A as its local vibration modes are in strong disagreement with experiment. However, configuration B agrees well with experiment and especially reproduces the local vibration E-line mode that is dominated by two substitutional C atoms and one interstitial Si atom between two C atoms and exhibits an asymmetric vibrational pattern for the two C atoms. Interestingly, our computations also suggest another local vibration mode at 101.6 meV (E-satellite) that is absent from experiment. Through a comparison of the potential-energy surfaces associated with both the E-line and E-satellite vibration modes, we observed the highly anharmonic phonon nature of the E-satellite mode, which would account for its large phonon linewidth and explain the absence of the corresponding peak in experimental observation. These findings suggest that caution should be taken in applying the harmonic approximation when comparing local vibration modes for defects in theory and experiment and calls for the development of a PL spectrum computation methodology that would formally include anharmonicity and phonon lifetime. Additionally, we demonstrate that carbon isotope experiments agree with our theoretical results. Our research not only provides insights into the PL spectra of the G center but also opens doors for the use of CS approaches to speed up the computation of PL spectra.

## ACKNOWLEDGEMENT

This work was supported by the U.S. Department of Energy, Office of Science, Basic Energy Sciences in Quantum Information Science under Award Number DE-SC0022289. This research used resources of the National Energy Research Scientific Computing Center, a DOE Office of Science User Facility supported by the Office of Science of the U.S. Department of Energy under Contract No. DE-AC02-05CH11231 using NERSC award BES-ERCAP0020966. C.L. acknowledges the support from the Sinergia project of the Swiss National Science Foundation (grant number CRSII5_189924).

## References

[1] G. Zhang, Y. Cheng, J. Chou, and A. Gali, Applied Physics Reviews **7** (2020).

[2] H. Bluhm, S. Foletti, I. Neder, M. Rudner, D. Mahalu, V. Umansky, and A. Yacoby, Nature Physics **7**, 109 (2011).

[3] J. W. Britton, B. C. Sawyer, A. C. Keith, C. J. Wang, J. K. Freericks, H. Uys, M. J. Biercuk, and J. J. Bollinger, Nature **484**, 489 (2012).

[4] S. Charpentier, L. Galletti, G. Kunakova, R. Arpaia, Y. Song, R. Baghdadi, S. M. Wang, A. Kalaboukhov, E. Olsson, and F. Tafuri, Nature communications **8**, 2019 (2017).

[5] A. Das, Y. Ronen, M. Heiblum, D. Mahalu, A. V. Kretinin, and H. Shtrikman, Nature communications **3**, 1165 (2012).

[6] E. Togan, Y. Chu, A. S. Trifonov, L. Jiang, J. Maze, L. Childress, M. G. Dutt, A. S. Sørensen, P. R. Hemmer, and A. S. Zibrov, Nature **466**, 730 (2010).

[7] A. L. Falk, P. V. Klimov, B. B. Buckley, V. Ivády, I. A. Abrikosov, G. Calusine, W. F. Koehl, Á Gali, and D. D. Awschalom, Phys. Rev. Lett. **112**, 187601 (2014).

[8] P. C. Humphreys, N. Kalb, J. P. Morits, R. N. Schouten, R. F. Vermeulen, D. J. Twitchen, M. Markham, and R. Hanson, Nature **558**, 268 (2018).

[9] C. T. Nguyen, D. D. Sukachev, M. K. Bhaskar, B. Machielse, D. S. Levonian, E. N. Knall, P. Stroganov, R. Riedinger, H. Park, and M. Lončar, Phys. Rev. Lett. **123**, 183602 (2019).

[10] T. Neuman, M. Eichenfield, M. E. Trusheim, L. Hackett, P. Narang, and D. Englund, npj Quantum Information **7**, 121 (2021).

[11] M. Pompili, S. L. Hermans, S. Baier, H. K. Beukers, P. C. Humphreys, R. N. Schouten, R. F. Vermeulen, M. J. Tiggelman, L. dos Santos Martins, and B. Dirkse, Science **372**, 259 (2021).

[12] G. Wolfowicz, F. J. Heremans, C. P. Anderson, S. Kanai, H. Seo, A. Gali, G. Galli, and D. D. Awschalom, Nature Reviews Materials **6**, 906 (2021).

[13] Y. Zhou, Z. Wang, A. Rasmita, S. Kim, A. Berhane, Z. Bodrog, G. Adamo, A. Gali, I. Aharonovich, and W. Gao, Science advances **4** (2018).

[14] Y. Xue, H. Wang, N. Xie, Q. Yang, F. Xu, B. Shen, J. Shi, D. Jiang, X. Dou, and T. Yu, The Journal of Physical Chemistry Letters **11**, 2689 (2020).

[15] H. Zeng, G. Duan, Y. Li, S. Yang, X. Xu, and W. Cai, Advanced functional materials **20**, 561 (2010).

[16] P. Udvarhelyi, B. Somogyi, G. Thiering, and A. Gali, Phys. Rev. Lett. **127**, 196402 (2021).

[17] T. T. Tran, K. Bray, M. J. Ford, M. Toth, and I. Aharonovich, Nature nanotechnology **11**, 37 (2016).

[18] Y. Jin, M. Govoni, G. Wolfowicz, S. E. Sullivan, F. J. Heremans, D. D. Awschalom, and G. Galli, Physical Review Materials **5**, 084603 (2021).

[19] A. Alkauskas, B. B. Buckley, D. D. Awschalom, and C. G. Van de Walle, New Journal of Physics **16**, 073026 (2014).

[20] L. Razinkovas, M. W. Doherty, N. B. Manson, C. G. Van de Walle, and A. Alkauskas, Physical Review B **104**, 045303 (2021).

[21] E. Londero, G. Thiering, L. Razinkovas, A. Gali, and A. Alkauskas, Physical Review B **98**, 035306 (2018).

[22] S. G. Cloutier, P. A. Kossyrev, and J. Xu, Nature materials **4**, 887 (2005).

[23] M. Hollenbach, N. Klingner, N. S. Jagtap, L. Bischoff, C. Fowley, U. Kentsch, G. Hlawacek, A. Erbe, N. V. Abrosimov, and M. Helm, Nature Communications **13**, 7683 (2022).

[24] W. Redjem, A. Durand, T. Herzig, A. Benali, S. Pezzagna, J. Meijer, A. Y. Kuznetsov, H. S. Nguyen, S. Cueff, and J. Gérard, Nature Electronics **3**, 738 (2020).

[25] W. Redjem, Y. Zhiyenbayev, W. Qarony, V. Ivanov, C. Papapanos, W. Liu, K. Jhuria, Z. Y. Al Balushi, S. Dhuey, and A. Schwartzberg, Nature Communications **14**, 3321 (2023).

[26] L. Komza, P. Samutpraphoot, M. Odeh, Y. Tang, M. Mathew, J. Chang, H. Song, M. Kim, Y. Xiong, and G. Hautier, arXiv preprint arXiv:2211.09305 (2022).

[27] M. Prabhu, C. Errando-Herranz, L. De Santis, I. Christen, C. Chen, C. Gerlach, and D. Englund, Nature Communications **14**, 2380 (2023).

[28] E. J. Candès and M. B. Wakin, IEEE Signal Process. Mag. **25**, 21 (2008).

[29] P. Hohenberg and W. Kohn, Physical review **136** (1964).

[30] G. Kresse and J. Furthmüller, Physical review B **54**, 11169 (1996).

[31] P. E. Blöchl, Physical review B **50**, 17953 (1994).

[32] J. P. Perdew, K. Burke, and M. Ernzerhof, Phys. Rev. Lett. **77**, 3865 (1996).

[33] J. P. Perdew, K. Burke, and Y. Wang, Physical review B **54**, 16533 (1996).

[34] J. Heyd, G. E. Scuseria, and M. Ernzerhof, J. Chem. Phys. **118**, 8207 (2003).

[35] A. Gali, E. Janzén, P. Deák, G. Kresse, and E. Kaxiras, Phys. Rev. Lett. **103**, 186404 (2009).

[36] D. Dhaliah, Y. Xiong, A. Sipahigil, S. M. Griffin, and G. Hautier, Physical Review Materials **6** (2022).

[37] F. Zhou, W. Nielson, Y. Xia, and V. Ozoliņš, Physical Review B **100**, 184308 (2019).

[38] K. Esfarjani and H. T. Stokes, Physical Review B **77**, 144112 (2008).

[39] A. Togo and I. Tanaka, Scr. Mater. **108**, 1 (2015).

[40] T. Tadano, Y. Gohda, and S. Tsuneyuki, Journal of Physics: Condensed Matter **26**, 225402 (2014).

[41] T. Tadano and S. Tsuneyuki, Physical Review B **92**, 054301 (2015).

[42] L. J. Nelson, G. L. Hart, F. Zhou, and V. Ozoliņš, Physical Review B **87**, 035125 (2013).

[43] A. Togo, L. Chaput, and I. Tanaka, Physical review B **91**, 094306 (2015).

[44] C. Lin, S. Poncé, and N. Marzari, npj Computational Materials **8**, 236 (2022).

[45] P. A. M. Dirac, Proceedings of the Royal Society of London.Series A, Containing Papers of a Mathematical and Physical Character **114**, 243 (1927).

[46] K. Huang and A. Rhys, Proceedings of the Royal Society of London.Series A.Mathematical and Physical Sciences **204**, 406 (1950).

[47] M. Lax, J. Chem. Phys. **20**, 1752 (1952).

[48] R. Kubo and Y. Toyozawa, Progress of Theoretical Physics **13**, 160 (1955).

[49] R. J. Bell, P. Dean, and D. C. Hibbins-Butler, Journal of Physics C: Solid State Physics **3**, 2111 (1970).

[50] S. A. Tawfik and S. P. Russo, Comput. Phys. Commun. **273**, 108222 (2022).

[51] W. M. Yim and R. J. Paff, J. Appl. Phys. **45**, 1456 (1974).

[52] W. Bludau, A. Onton, and W. Heinke, J. Appl. Phys. **45**, 1846 (1974).

[53] G. E. Jellison Jr, J. Appl. Phys. **53**, 5715 (1982).

[54] L. W. Song, X. D. Zhan, B. W. Benson, and G. D. Watkins, Physical Review B **42**, 5765 (1990).

[55] C. Linderälv, W. Wieczorek, and P. Erhart, Physical Review B **103**, 115421 (2021).

[56] R. O. Jones and O. Gunnarsson, Reviews of Modern Physics **61**, 689 (1989).

[57] A. Gali and J. R. Maze, Physical Review B **88**, 235205 (2013).

[58] Y. Xiong, C. Bourgois, N. Sheremetyeva, W. Chen, D. Dahliah, H. Song, J. Zheng, S. M. Griffin, A. Sipahigil, and G. Hautier, Science Advances **9** (2023).

[59] V. Ivanov, J. Simoni, Y. Lee, W. Liu, K. Jhuria, W. Redjem, Y. Zhiyenbayev, C. Papapanos, W. Qarony, and B. Kanté, Physical Review B **106**, 134107 (2022).

[60] A. R. Bean, R. C. Newman, and R. S. Smith, Journal of Physics and Chemistry of Solids **31**, 739 (1970).

[61] K. Thonke, H. Klemisch, J. Weber, and R. Sauer, Physical Review B **24**, 5874 (1981).

[62] C. Beaufils, W. Redjem, E. Rousseau, V. Jacques, A. Y. Kuznetsov, C. Raynaud, C. Voisin, A. Benali, T. Herzig, and S. Pezzagna, Physical Review B **97**, 035303 (2018).

[63] K. Esfarjani and H. T. Stokes, Physical Review B **77**, 144112 (2008).

[64] J. Zheng, D. Shi, Y. Yang, C. Lin, H. Huang, R. Guo, and B. Huang, Physical Review B **105**, 224303 (2022).

[65] See the Supplemental Material at xxx for parameters containing harmonic phonon propertie and other information. (unpublished).

[66] K. Murata, Y. Yasutake, K. Nittoh, S. Fukatsu, and K. Miki, Aip Advances **1** (2011).

[67] P. Han and G. Bester, Physical Review B **106**, 245404 (2022).

[68] J. Hafner and M. Krajci, Journal of Physics: Condensed Matter **5**, 2489 (1993).

[69] T. Tadano and S. Tsuneyuki, Phys. Rev. Lett. **120**, 105901 (2018).

[70] F. Bertoldo, S. Ali, S. Manti, and K. S. Thygesen, npj Computational Materials **8**, 56 (2022).

[71] J. Davidsson, W. Stenlund, A. S. Parackal, R. Armiento, and I. A. Abrikosov, arXiv preprint arXiv:2306.11116 (2023).